%% file: hypermuchfuss_aa1.tex
\documentclass{aa}
\usepackage{graphicx} \usepackage{natbib} \usepackage{booktabs}
\bibpunct{(}{)}{;}{a}{}{,} \newcommand{\hei}{He~{\sc i}}

 \newcommand{\teff}{$T_{\rm eff}$}
 
 \newcommand{\msun}{$\rm{M_\odot}$}
\newcommand{\kms}{$\rm{km~s^{-1}}$} \newcommand{\logg}{$\log{g}$}
\newcommand{\logn}{$\log{n_{\rm{He}}/n_{\rm{H}}}$} 
\newcommand{\pmp}{$\pm$} \newcommand{\Fe}{$[Fe/H]$}

\begin{document}


\hyphenation{ana-lysed do-mi-nant ana-ly-sis}

\title{The Hyper-MUCHFUSS project: probing the Galactic halo with sdB stars\thanks{Based on data from the Sloan Digital Sky Survey, and data collected at the 3.5m telescope at DSAZ observatory (Calar Alto) in Spain, the 4.2m William Herschel Telescope on La Palma, and the European Southern Observatory in Chile according to the programs 081.D-0819(A),082.D-0649(A) and 084.D-0348(A).}}
\author{A.~Tillich\inst{1}\and U.~Heber\inst{1}\and S.~Geier\inst{1}\and H.~Hirsch\inst{1}\and P.~F.~L.~Maxted\inst{2}\and B.~T.~G\"ansicke\inst{3}\and T.~R.~Marsh\inst{3}\and R.~Napiwotzki\inst{4}\and R.~H.~\O{}stensen\inst{5}\and R.-D.~Scholz\inst{6}}
\offprints{A.~Tillich
\email{Alfred.Tillich@sternwarte.uni-erlangen.de}}
\institute{Dr. Karl Remeis-Observatory \& ECAP, Astronomical Institute, Friedrich-Alexander University Erlangen-Nuremberg, 
Sternwartstr.7, 96049 Bamberg, Germany\and Astrophysics Group, School of Physics and Geographical Sciences, Lennard-Jones Laboratories, Keele University, ST5 5BG, United Kingdom\and Department of Physics, University of Warwick, Coventry CV4 7AL, United Kingdom\and Centre of Astrophysics Research, University of Hertfordshire, College Lane, Hatfield AL10 9AB, United Kingdom\and Instituut voor Sterrenkunde, K.U. Leuven, 
Celestijnenlaan 200D, 3001 Leuven, Belgium\and Astrophysikalisches Institut Potsdam, An der Sternwarte 16, 
D-14482 Potsdam}

\date{Received 2010 / Accepted 2010}

\abstract{
High-velocity stars in the Galactic halo, e.g. the so-called hyper-velocity stars (HVS), are important tracers of the properties of the dark matter halo, in particular its mass. 
}{
A search for the fastest stars among hot subdwarfs (sdB) in the halo is carried out to identify HVS, unbound to the Galaxy, and bound population II stars in order to derive a lower limit to the halo mass. 
}{
Based on the SDSS DR6 spectral database we selected stars with high rest-frame velocities. These radial velocity measurements were verified at several telescopes to exclude radial velocity variable stars. 
Out of 88 stars observed in the follow-up campaign 39 stars were found to have constant radial velocities. For twelve of them we measured a proper motion significantly different from zero and obtained spectroscopic distances from quantitative spectral analysis to construct the full 6D phase space information for a kinematical study.
}{
All but one programme sdBs show halo characteristics, but can be distinguished into two kinematical groups, one (G1) with low Galactic rotation typical of halo stars and a second one (G2) with rapid retrograde motion. 
We also investigate the possibility that the programme stars are not genuine halo stars but ejected from the Galactic disc or bulge. 
The G1 objects crossed the Galactic plane in the central bulge, whereas the G2 stars did in the outer Galactic disc. 
J1211+1437 (G2) is a HVS candidate, as it is unbound to the Galaxy if the standard Galactic potential is adopted. 
}{
We conclude that in the ejection scenario G1 stars might have been formed via 
the slingshot mechanism that invokes acceleration by tidal interaction of a binary with the central supermassive black hole. 
The G2 stars, however, would originate in the outskirts of the Galactic disc and not in the central bulge. 
J1211+1437 is the first unbound subdwarf B star, for which we can rule out the slingshot mechanism. 
Alternatively, we may assume that the stars are old population II stars and therefore have to be bound. Then the kinematics of J1211+1437 set a lower limit of $2\times 10^{12}$ M$_\odot$ to the mass of the Galactic dark matter halo. 
}


\maketitle

\section{Introduction}\label{sec:intro}

The properties of the dark matter halo are important
to understand how the Galaxy formed and evolved.
Observations of halo stars put constraints on  
theoretical models of halo formation \citep[e.g.][]{1996ApJ...462..563N}. Large surveys, such as the Sloan Digital Sky Survey
\citep[SDSS,][]{2000AJ....120.1579Y} and the RAdial Velocity Experiment
\citep[\emph{RAVE},][]{2006AJ....132.1645S}, 
provide large numbers of stars to trace the halo properties, such as the total mass of the halo.

Globular clusters, satellite 
galaxies, as well as large samples of halo stars, respectively, have been used to estimate the halo mass. Actually only the objects with the most extreme velocities 
provide tight constraints and, hence, the mass estimates depend mostly on them \citep{2003A&A...397..899S,2007MNRAS.379..755S}.
A halo mass of about $2\times10^{12}$\,\msun\ 
was favoured in earlier investigations
\citep{1999MNRAS.310..645W,2003A&A...397..899S}, while more recent studies
prefer lower masses of about half that value 
\citep{2005MNRAS.364..433B,2007MNRAS.379..755S,2008ApJ...684.1143X}. 

The hyper-velocity stars \citep[HVS, ][]{2005ApJ...622L..33B,2005A&A...444L..61H,2005ApJ...634L.181E} 
are the fastest moving stars known in the halo. 
 Their supposed
place of origin is the Galactic centre, where they have been suggested to be
accelerated by tidal interactions of a binary star with the super-massive black
hole \citep[SMBH, ][]{1988Natur.331..687H}.
Whether a HVS can in fact escape from the Galaxy or not depends on the halo mass \citep{2009ApJ...691L..63A}. 

Kinematical studies of the hyper-velocity stars were based on their radial velocities only. Recently, \cite{2009A&A...507L..37T} were able to 
measure proper motions of an A-type HVS and study its 3-D kinematics to trace its place of birth in the Galactic disc. They found it to originate far from the Galactic centre, thereby challenging the SMBH-slingshot mechanism of \cite{1988Natur.331..687H}. Hence \citet{2009A&A...507L..37T} suggested a runaway mechanism for 
the star's formation. Further evidence that such a mechanism works comes from two similar studies of the hyper run-away stars HD 217791 \citep{2008A&A...483L..21H} and HIP~60350 \citep{2010ApJ...711..138I}, which were also found to originate in the outer rim of the Galactic disc nowhere near the Galactic centre. 

While most of the 17 HVS known today \citep{2009ApJ...690.1639B, 2009A&A...507L..37T} are early-type main-sequence stars, there is just one evolved low-mass star, US~708, a hot subdwarf star of spectral type sdO \citep{2005A&A...444L..61H}. 

Most of the previous studies of halo stars to constrain the dark matter properties are hampered by the lack of proper motion measurements. Hence they had to rely substantially on radial velocity distributions.
In such cases only four coordinates (i.e. two position values,
distance and radial velocity, RV) of the 6D phase space are
determined and the missing proper motion components are
handled in a statistical approach.
In the presently most extensive study \citet{2008ApJ...684.1143X} measured radial velocities for more than 10,000 blue halo stars from the SDSS and classified their sample as a mix of blue horizontal branch (BHB) stars, blue stragglers and main-sequence stars with effective temperatures roughly between 7,000 and 
10,000\,K according to their colours. \citet{2008ApJ...684.1143X} selected 2400 blue horizontal-branch stars 
to estimate the halo mass out to 60~kpc to be $1.0\times10^{12}$ M$_\odot$ using a halo model of 
\cite{1997ApJ...490..493N}. For one star from that sample \cite{2010ApJ...718...37P} were able to obtain proper motion and carry out a detailed kinematic analysis, which revealed an inbound 
Population~II horizontal branch star with a
Galactic rest-frame (GRF) velocity of $\sim$700\,\kms\ at its current
position. This makes it the fastest halo star known, and provided 
a lower limit of $1.7\times10^{12}$ M$_\odot$ for the total halo mass of the Galaxy, significantly 
exceeding the value determined by \citet{2008ApJ...684.1143X}. 

This example shows that it is rewarding to study the kinematics of additional stars in the halo and to consider classes of stars other than BHB stars, as well. Of course, the Galactic halo hosts a plethora of white dwarfs \citep{2006ApJS..167...40E}. However, they are so faint that they can be analysed in the solar neighbourhood only. 
Another group of evolved low mass stars are the hot subdwarf stars (sdB, sdO) that dominate the population of faint blue stars at high Galactic latitudes to visual magnitudes of about V=18 \citep{1986ApJS...61..305G}. 
They are considered to be helium core burning stars with very thin ($<$0.02\,$M_\odot$) inert 
hydrogen envelopes and masses around 0.5~\msun. Following ideas 
outlined by \cite{1986A&A...155...33H}, the sdBs
can be identified with models for extreme horizontal branch (EHB) stars. An
EHB star bears great resemblance to a helium main-sequence star of half a
solar mass and it should evolve similarly, i.e. directly to the white dwarf
cooling sequence, bypassing a second giant phase \citep[for a review see][]{2009ARA&A..47..211H}. 
For the formation of subdwarf B stars three scenarios are discussed by \cite{2003MNRAS.341..669H}: 
common envelope ejection, stable Roche lobe overflow (RLOF), and the merger of two helium white dwarfs. 
Some alternate scenarios for the formation of single sdB stars are reviewed by \cite{2009CoAst.159...75O}.

Hot subdwarf stars exist in the field of the Galaxy but also in globular clusters, in the Galactic bulge and have even been resolved in the elliptical galaxy M~32 \citep{2008ApJ...682..319B}. Kinematical studies \citep{2004A&A...414..181A, 2008ASPC..391..257N} indicate that they occur in all stellar populations of the Galaxy. 

However, very little is known about the halo population of hot subdwarfs except those in globular clusters. Some high-velocity hot subdwarfs have attracted interest because of their high radial velocities, most notably, the sdO star US~708, whose radial velocity in the rest-frame was measured at 751 \kms \citep{2005A&A...444L..61H} -- the second HVS star discovered. Unfortunately we cannot deduce the 
origin of the star, as we lack a reliable proper motion measurement. 

Motivated by the discovery of US~708, we embarked on a project to identify a sample of population II hot subdwarfs and study their kinematics from radial velocity and proper motion.
We make use of the MUCHFUSS survey \citep{geier_TARGET}, which searches for 
close binaries with high radial-velocity variations. The search strategy also provides targets that are 
not close binaries but travel through space at high RV without variations. 
These stars are the targets of 
our investigation. Accordingly we entitled our project Hyper-MUCHFUSS as we provide an extension of MUCHFUSS. 

The paper is organised as follows. In Sect.~\ref{sec:survey} we introduce our 
survey for HVS and in Sect.~\ref{sec:PM} our sophisticated proper 
motion measurement method. The kinematical analysis techniques 
are shown in Sect.~\ref{sec:DIST_KINE}. In Sect.~\ref{sec:results} we 
present our results and summarise and conclude in Sect.~\ref{sec:conclu}.

\section{Survey}\label{sec:survey}
\begin{figure}[t!]
\begin{center}
\includegraphics[scale=0.4]{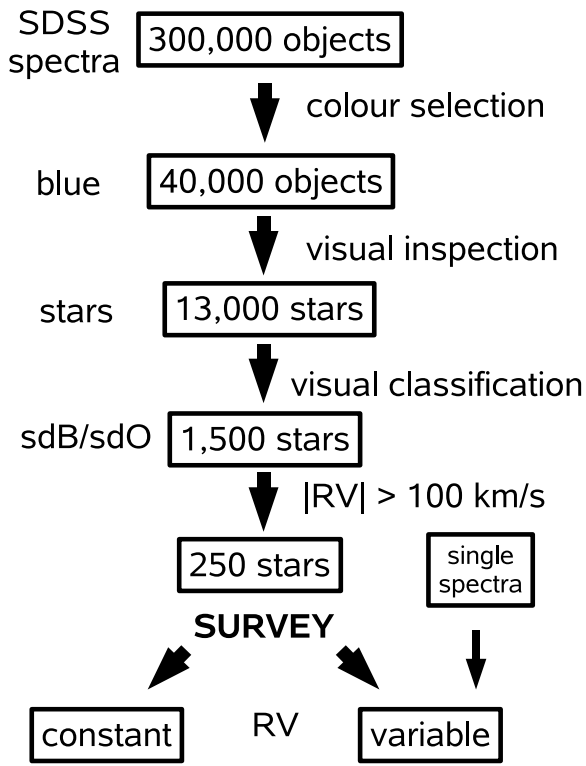}
 \caption{Target selection method based on SDSS. Stars are selected for which the absolute value of 
the RV exceeds 100~\kms. Hence disc stars are efficiently removed.} 
 \label{fig:fluss}
\end{center}
 \end{figure}

The enormous SDSS database is well evaluated in terms of errors and accuracy. Hence it is the perfect starting point for the MUCHFUSS survey. In order to select subdwarf candidates, we used several indicators, e.g. colour, spectral classification and radial velocity (RV). Fig.~\ref{fig:fluss} shows a flowchart of the target selection method. We selected sdO/B candidates 
by colour \citep[g-r$<$0.1 and u-g$<$0.4, see][]{geier_TARGET} 
and pre-classified their spectra by visual inspection. 
Measuring the radial velocity by fitting synthetic models, we selected only stars faster than $\pm$100 \kms. 
For most of the known sdB binaries, the RV semi-amplitudes are below 100~\kms \citep{geier_TARGET}. 
Hence the radial velocity of a typical sdB binary of the Galactic disc will rarely exceed 100~\kms\ in absolute value 
and these stars are consequently excluded in MUCHFUSS. 
We converted the heliocentric RV to the Galactic rest-frame (GRF). 
The larger the RV the higher is the priority we assign to the target for our survey. 
Especially stars with absolute GRF velocities of more 
than 275 \kms\ are high priority HVS candidates. This observational cut was 
introduced by \cite{2007ApJ...660..311B} to distinguish HVS from halo stars by 
their origin. More information on MUCHFUSS and the process of target selection 
is presented by \cite{geier_TARGET}.

More than 250 targets with measured radial velocity remained on the MUCHFUSS target list, serving as a 
first epoch. Second-epoch spectra were obtained with ESO-VLT/FORS ($R \approx 1800, \lambda \approx
3730-5200$\AA{}), WHT/ISIS ($R \approx 1800, \lambda \approx 3730-5200$\AA{}), CAHA-3.5m/TWIN 
($R \approx 4000, \lambda \approx 3460-5630$\AA{}) and ESO-NTT/EFOSC2 ($R \approx 2200, \lambda \approx 4450-5110$\AA{}). 
Due to the different wavelength coverage and resolution, the number of visible absorption lines is 
correspondingly restricted\footnote{Note that the EFOSC2 spectra cover only H$_\beta$ and 2 \hei\ lines.}. 
Up to now, such spectra are available for 88 stars, which is about one third of the target list. 
The RVs have been measured by $\chi^2$-fitting of suitable synthetic spectra. Based on the 
MUCHFUSS RV list, we regard a star as RV constant if its velocity is consistent with the first 
epoch within the respective error limits. However it is obvious that these errors individually depend 
on a variety of conditions like e.g. the S/N (see Table \ref{tab_RV}), the resolution, wavelength coverage and the number of visible absorption lines. 

\begin{table*}
\caption{Heliocentric radial velocity table of the target sample with SNR per pixel in the continuum next to H$_\beta$.\label{tab_RV}}
\begin{center}
\begin{tabular}{llllllc}
\midrule\midrule
name & short name & type & OBS & MJD & $v_{rad}$ & SNR@H$_\beta$ \\
     &     &     &     &     & \kms   &  \\
\midrule
\input{tab_RV2.list}
\midrule
\end{tabular}
\end{center}
\end{table*}

The MUCHFUSS results for the binaries will be presented in a dedicated paper by \citet[][in prep.]{geier_HMF}. 
While 49 of the target stars were found to be RV variable, 39 do not show variations and therefore 
make up the sample that is studied in this paper. 
In the following sections we will exclusively focus on these candidates. 

 \begin{figure}[h!]
\begin{center}
      \resizebox{\hsize}{!}{\includegraphics{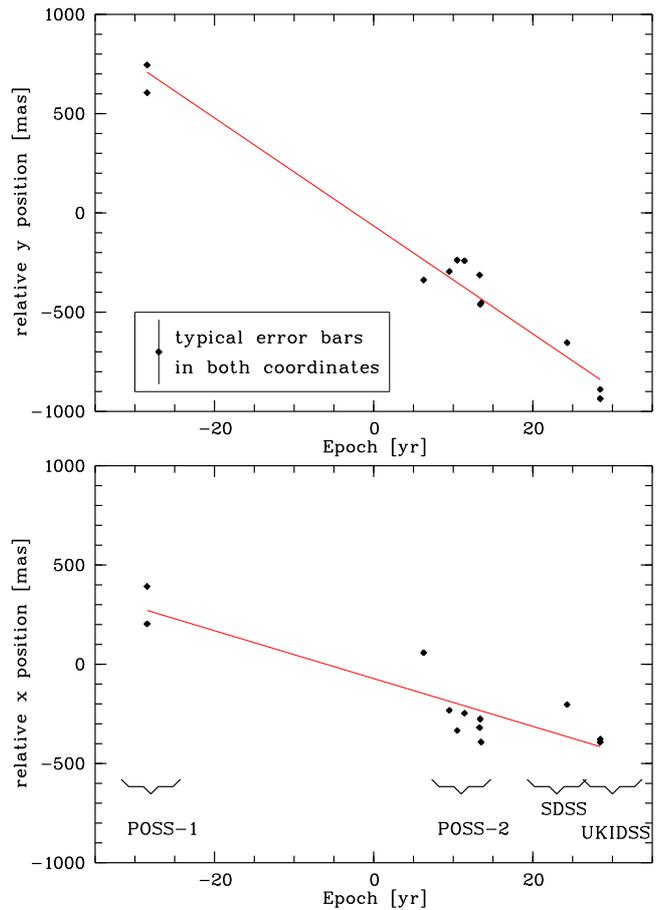}}
 \caption{Proper motion components derived from the position measurements for 
 J1211+1437, where 1978.77 is the zero epoch.} 
 \label{fig:PMfit_HMF}
\end{center}
 \end{figure}

\section{Proper motion}\label{sec:PM}
A kinematical analysis can only be done, if the star's location and space velocity is known. 
Hence we attempt to measure proper motions and determine spectroscopic distances for all our 39 
candidates. 
We collected all available independent position measurements on Schmidt plates
(APM - \citealt{2000yCat.1267....0M};
SSS - \citealt{2001MNRAS.326.1279H}) 
and combined them with the SDSS and other available positions
(CMC14~\citealt{2006yCat.1304....0C};
2MASS - \citealt{2003tmc..book.....C};
UKIDSS - \citealt{2007MNRAS.379.1599L}). 
Multi-epoch SDSS astrometric measurements, 
taken from the Princeton SDSS database \citep{2004AJ....128.2577F} 
were included in the proper motion fit. 
We obtained more measurements of Schmidt plates, 
from up to 14 different epochs in case
of overlapping plates of the Digitised Sky
Survey\footnote{http://archive.stsci.edu/cgi-bin/dss\_plate\_finder} (DSS).
The FITS images of 15 by 15 arcmin size were extracted from all available plates 
and the ESO MIDAS tool center/gauss was used to measure positions. 
To measure absolute proper motions, we initially had to find and identify compact background galaxies 
from the SDSS. It is highly important to obtain enough galaxies well distributed 
over the image, although galaxies tend to cluster. 
We used at least 10 galaxies per measurement and determined 
the position of the star relative to the reference galaxies. Finally we transformed 
the target positions on all the Schmidt plates to the SDSS system. The small fields 
allowed us to apply a simple model (shift+rotation). 
The typical error in both coordinates 
at one epoch, including the error of the transformation using the measured 
galaxies and the individual centroiding error for the target, was 150mas. 
We obtained one position 
per epoch and used linear regression to derive the proper motions with their errors 
(see e.g. Fig~\ref{fig:PMfit_HMF}). 

\begin{table}
\caption{PM measurements for the positive detections together with the number of epochs n$_{ep}$ and the number of reference galaxies n$_{gal}$. The V magnitudes have been derived from SDSS photometry using the transformations of \cite{2006A&A...460..339J}. \label{tab_PM}}
\begin{center}
\begin{tabular}{lllll}
\midrule\midrule
name & V & $(\mu_\alpha\cos(\delta))$ & $\mu_\delta$ & n$_{ep}$/n$_{gal}$ \\
     & mag & mas/yr  & mas/yr       &      \\
\midrule
\input{tab_PMok.dat}
\midrule
\end{tabular}
\end{center}
\end{table}

Especially for the old POSS-I epoch we noticed large discrepancies between the positions, 
which are probably due to the different colour filters and quality of the plates. 
The best way to minimize possible systematic effects is simply to use the same 
set of reference galaxies for every epoch. We regard a proper motion as detected, if the 
position measurements do not show a large spread relative to the linear fit and the derived proper 
motion is significantly different from zero. 
27 stars turned out to have a proper motion consistent with zero, while for 12 of our 39 candidates 
($\approx 38\%$) a significant proper motion was measured. 
One of them is the sdB J2156+0036 which has already been used to show the 
potential of our method \citep[see Fig.~1 of][]{2010Ap&SS.tmp..117T}. This star represents 
also a typical star with proper motion from our target sample. 
Note that for all but two objects (J1644+4523 and J0849+1455)
the determined proper motions agree within their
errors with those of the recently published
PPMXL catalogue \citep{2010AJ....139.2440R},
where for all objects our proper motion errors are smaller than
the PPMXL ones. 

Tables \ref{tab_RV} and \ref{tab_PM} list the measurements for the sample analysed in this paper.
In Fig~\ref{fig:PMfit_HMF} we demonstrate the measurement of the proper motion for J1211+1437, an outstanding target that will be discussed in detail in section \ref{sec:j1211}.  
 

\section{Distance and Kinematics}\label{sec:DIST_KINE}
The second important parameter is the distance, necessary to determine the current location of the stars 
and their transversal velocities along with their proper motions. 
We determine the distance from a quantitative spectral analyse, which provides the effective temperature, gravity and helium abundance. The distance is derived from the atmospheric parameters and the apparent magnitude, by adopting the canonical mass of 0.48~\msun. 

In most of the cases the stellar spectra from the SDSS provide sufficient S/N and resolution for 
a quantitative spectral analysis. Furthermore, in some cases they make up the most reliable data we have. 
We applied $\chi^2$-fitting of synthetic line profiles to the 
Balmer and helium lines in order to determine the atmospheric parameters and abundances 
(see Fig. \ref{fig:J1644+J1211_spec}). 
Inevitably, our selection procedure produces some misclassifications, therefore we also ended up with two DA white dwarfs in our sample. 
All of the subdwarfs have been analysed using fully metal line-blanketed LTE models of solar metallicity 
\citep {2000A&A...363..198H}\footnote{For the DA white dwarfs we used the synthetic models by \cite{2009A&A...498..517K}.}. 
Finally spectroscopic distances are calculated using the astrophysical fluxes following 
\cite{2001A&A...378..907R}. 
Independent studies showed, that the atmospheric parameters (and hence the distances) depend only little on the choice of the metallicity. \citet{2000A&A...363..198H} studied the effect by comparing results from models with solar composition to those of 1/100 solar metallicity in a detailed study of high resolution spectra for three sdB stars. The differences are of the order of 200-300~K in \teff\ and ~0.03~dex in \logg, far lower than the uncertainties of our results. Hence, metallicity effects are not significant. 
The uncertainties of the distance have been determined using a Monte Carlo approach for 
the input quantities \citep[cf. ][]{2009A&A...507L..37T,2010A&A...517A..36T,2010ApJ...718...37P}. 

Applying the Galactic potential of \citet{1991RMxAA..22..255A} we calculated orbits and 
reconstructed the path of the star with the program of \citet{1992AN....313...69O}. 
The distance of the GC from the Sun was adopted to be 8.0~kpc and
the Sun's motion with respect to the local standard of rest was taken from \citet{1998MNRAS.298..387D}. 
Based on our data we are not able to draw conclusions on detailed structural properties of different 
galactic potentials. However, we are able to discuss global effects (see Section \ref{sec:results}). 
The error in space motion is dominated by that of the distance (via the 
gravity error) and those of the proper motion components\footnote{Although some of the second-epoch spectra are of low quality, the RV errors (see Table~\ref{tab_RV}) are irrelevant for the error budget. 
The distribution of the space velocity is dominated by the errors in the transversal velocity components, which may have uncertainties as large as $\pm100$~\kms.}. 
Varying these three quantities within their respective errors 
we applied a Monte Carlo procedure for the error propagation and to derive the median GRF 
velocities at the present location and their distribution \citep[see Fig.~\ref{fig:J1211_SHIST}, cf. ][]{2009A&A...507L..37T,2010A&A...517A..36T,2010ApJ...718...37P}. 
We compared our results with the local escape velocities as calculated 
from the Galactic potential of \citet{1991RMxAA..22..255A}. 
Furthermore, from kinematic characteristics (U, V, eccentricity \textit{e} and the z-component of 
the angular momentum J$_Z$) we 
obtained additional information about the population membership of the stars. Following the trajectories backwards in time we can even put constraints on the origin of the stars, applying the same Monte Carlo method. 

\begin{figure}[t!]
\begin{center}
\includegraphics[angle=0,scale=0.7]{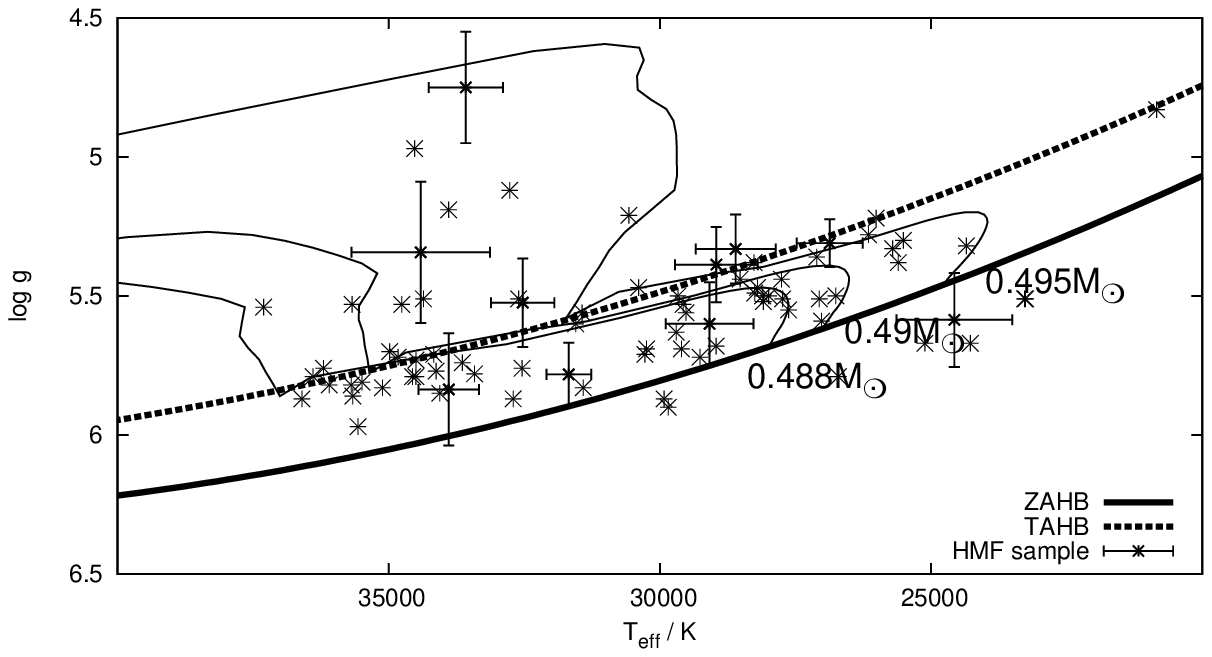}
\caption[\teff-\logg-diagram for the HMF subdwarf B stars]{Comparison of the position of the 10 subdwarfs 
(with error bars) from HMF project in the \teff-\logg-diagram to evolutionary tracks of 
\citet{1993ApJ...419..596D} for a metallicity of \Fe$=-1.48$. The ZAHB and the TAHB are indicated 
only from fitting of the tracks. 
The SPY subdwarf B sample is plotted for reference \citep{2005A&A...430..223L}. 
Note that two subdwarfs lie lie well above the canonical TAHB and therefore most likely are in a 
shell-He burning post-EHB stage of evolution.
}
\label{fig:Etracks_HMFsdB}
\end{center}
\end{figure}

\section{Results}\label{sec:results}

Until now about one third of the targets of the MUCHFUSS survey have been observed. 
Accordingly 39 of them have constant RV within the detection limits. 
Twelve stars showed proper motions significantly different from zero and have been analysed in 
detail using quantitative spectroscopy. In Fig.~\ref{fig:J1644+J1211_spec} we show a comparison between the observed 
and the synthetic spectra for two subdwarf stars. 
Two stars turned out to be DA white dwarfs rather than sdB stars. 
The derived stellar parameters are shown in Table~\ref{tab_param}, 
while the respective RV can be found in Table~\ref{tab_dist}.
\begin{table}
\begin{small}
\caption{Stellar parameters for the remaining stars with PM significantly different from zero.\label{tab_param}}
\begin{center}
\begin{tabular}{lcccc}
\midrule\midrule
name & type & \teff & \logg & \logn \\
     &  & K  &  &      \\
\midrule
\input{PM_ok_paramtex.tab_RD}
\midrule
\end{tabular}
\end{center}
\end{small}
\end{table}

In Fig.~\ref{fig:Etracks_HMFsdB} we show a \teff-\logg-diagram for the 10 subdwarfs in comparison to the reference sample of the SPY sdB stars analysed by \cite{2005A&A...430..223L}. 
Obviously 8 of our subdwarf B stars reside on the extended horizontal-branch (EHB) within the respective errors. 
However, the stars J1556+4708 and J2244+0106 lie well above the EHB, and must therefore be in a post-EHB stage of evolution if they have the canonical mass of 0.48~\msun. 
Fig.~\ref{fig:Etracks_HMFsdB} shows that our subdwarf B sample goes well together with the SPY sample. 
Based on these parameters we determined the distances, which are given in Table~\ref{tab_dist}. 
Together with the median space velocity we also obtained 
the escape velocity in the Galactic potential of \cite{1991RMxAA..22..255A}. 
\begin{table}
\caption{Estimated distance, escape and GRF velocities (median) and local escape velocities for the Galactic potential of \cite{1991RMxAA..22..255A}.  \label{tab_dist}}
\begin{center}
\begin{tabular}{lcccr}
\midrule\midrule
name & dist & $v_{rad}$ & $v_{GRF}$ & $v_{esc}$  \\
 & kpc & \kms & \kms & \kms \\
\midrule
\input{tab_vGRF.list}
\midrule
\end{tabular}
\end{center}
\end{table}
To further quantify the kinematics of the stars, we made a comparison with the 
kinematics of white dwarfs. A sample of 398 white dwarfs from the SPY survey was 
studied by \cite{2006A&A...447..173P}. Based on the 3D-orbit, the $V-U$ diagram and the 
\textit{e}-$J_{\rm Z}$-diagram they introduced a kinematic population classification scheme 
and combined it with age information. They performed a 
detailed kinematical analysis accounting for errors by means of a Monte Carlo error 
propagation code, similar to our method. A sample of abundance selected typical main-sequence 
stars served as reference sample. They derived the 3$\sigma$-contours for the $V-U$ diagram 
on which the kinematic classification is based. For the \textit{e}-$J_{\rm Z}$ diagram 
a ``Region B'' is defined 
such that it excludes as many thick-disc stars as possible. 
The last criterion is the $\rho$-Z-diagram, which is used to classify the orbits among the populations 
by comparing with template Galactic orbits \citep{2003A&A...400..877P}. 
A substantial thick disc fraction of 7\% was found, while only 2\% of the DAs show characteristic 
halo properties. 

\begin{figure*}[t!]
\includegraphics[scale=0.72]{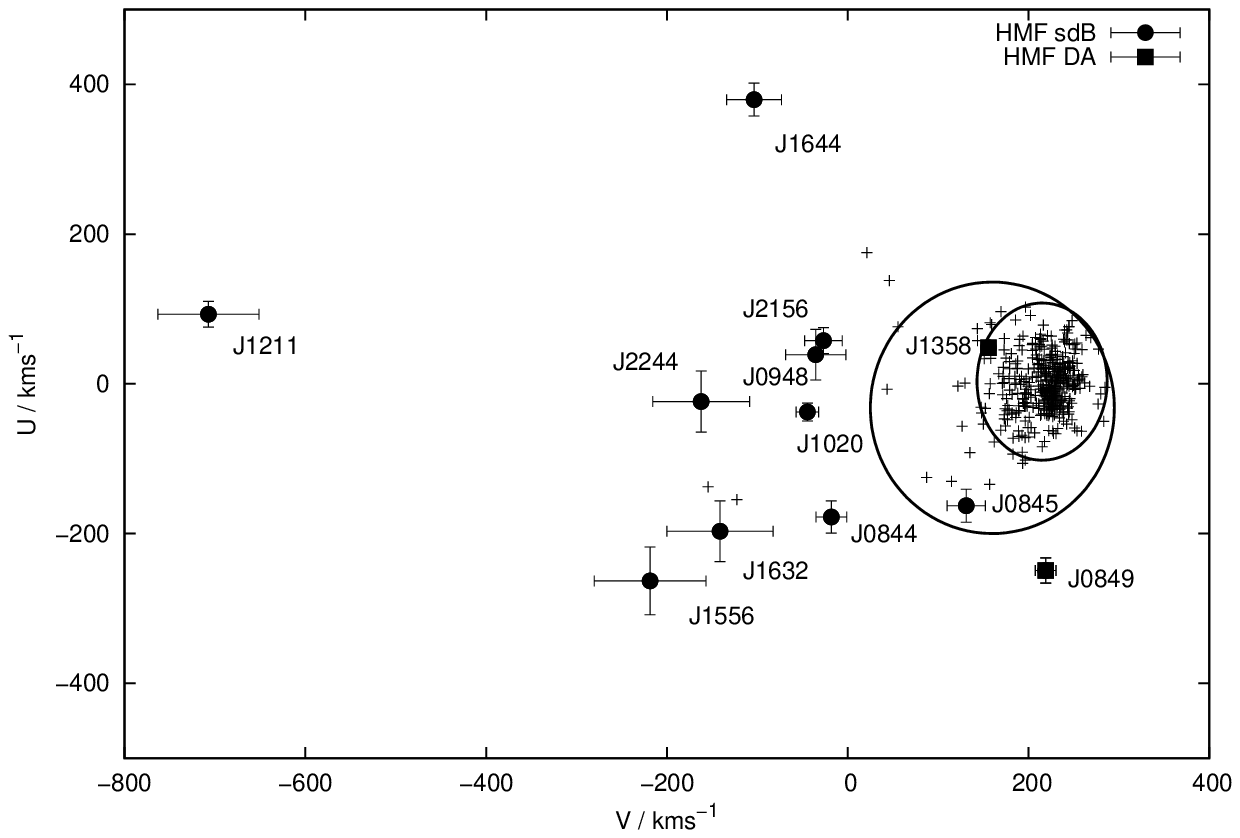}
\includegraphics[scale=0.72]{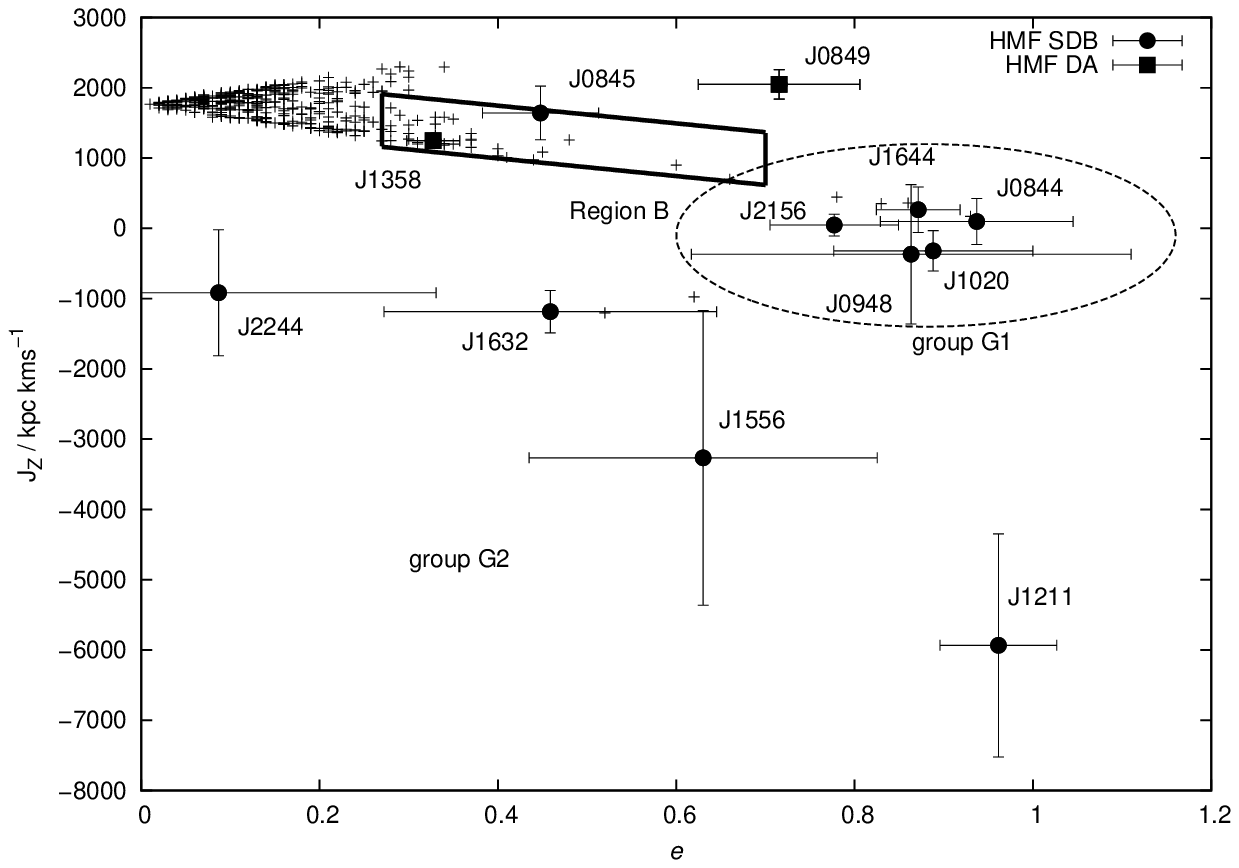}
\caption{
V-U (\textit{left}) and \textit{e}-J$_Z$ diagram (\textit{right}) for our 12 targets. 
The white dwarf 
sample (+) of \cite{2006A&A...447..173P} serves as reference. The solid ellipses render the $3\sigma$-thin and thick disc 
contours, while the solid box marks the thick disc region (Region B). 
Note that 10 of our targets are outside of the regions for the two disc populations in both diagrams, but form two 
distinct groups. The 5 stars inside the dashed ellipse are called group 1 (G1), while the stars with negative J$_Z$ are defined as group 2 (G2).
}
\label{fig:kine}
\end{figure*}

\begin{figure*}
\includegraphics[scale=0.72]{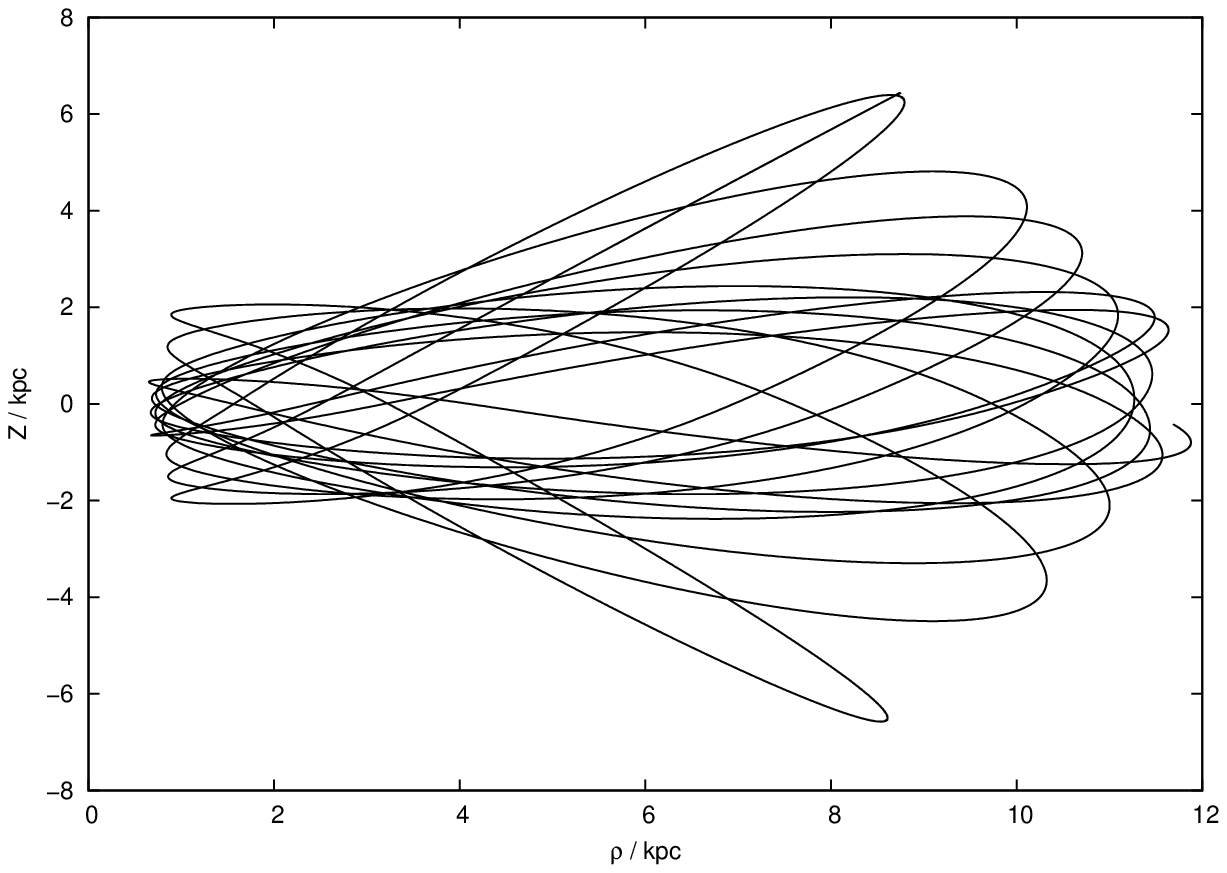}
\includegraphics[scale=0.72]{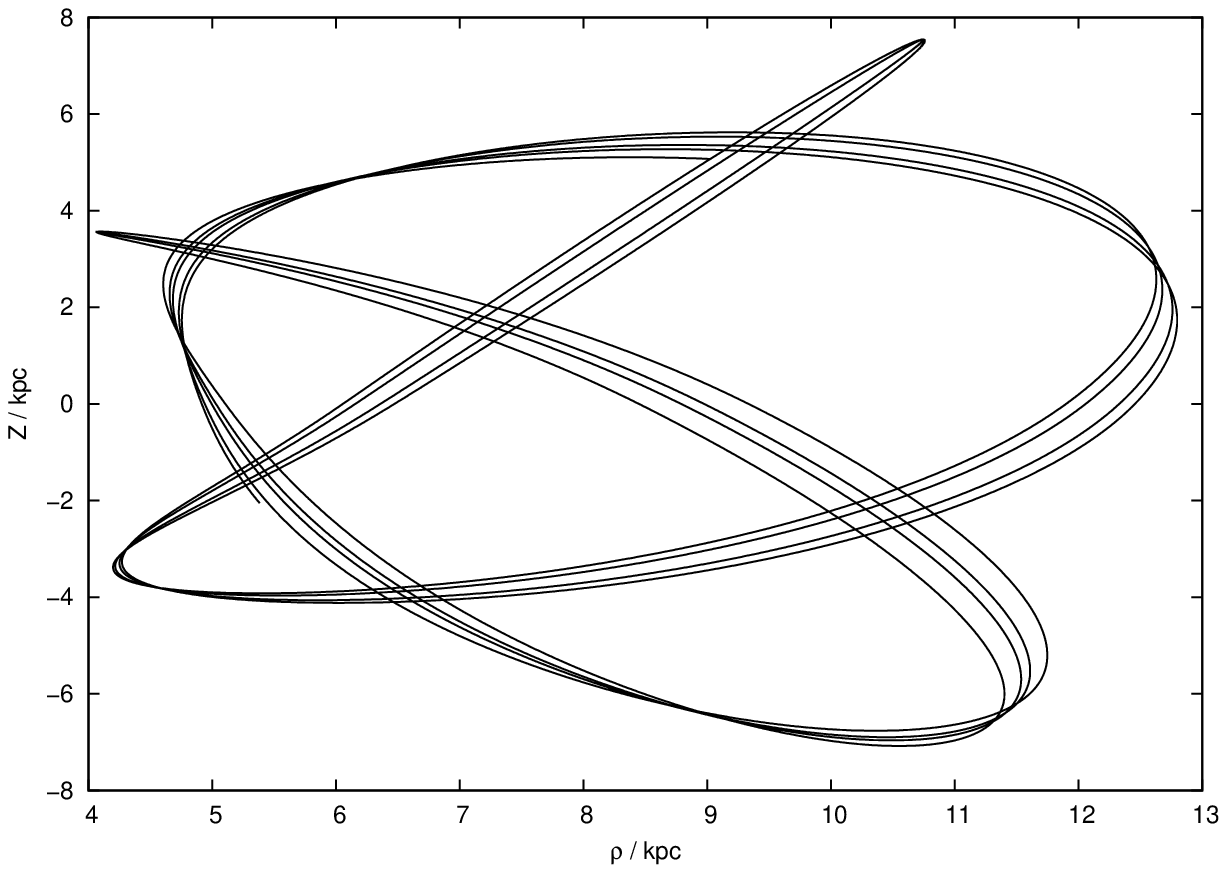}
\caption{   
$\rho$-Z-diagrams for the two group representatives, J1020+0137 (group 1, \textit{left}) and 
J1632+2051 (group 2, \textit{right}).
}
\label{fig:kine2}
\end{figure*}

\begin{figure*}[tb]
\begin{center}
\includegraphics[angle=0,scale=1.0]{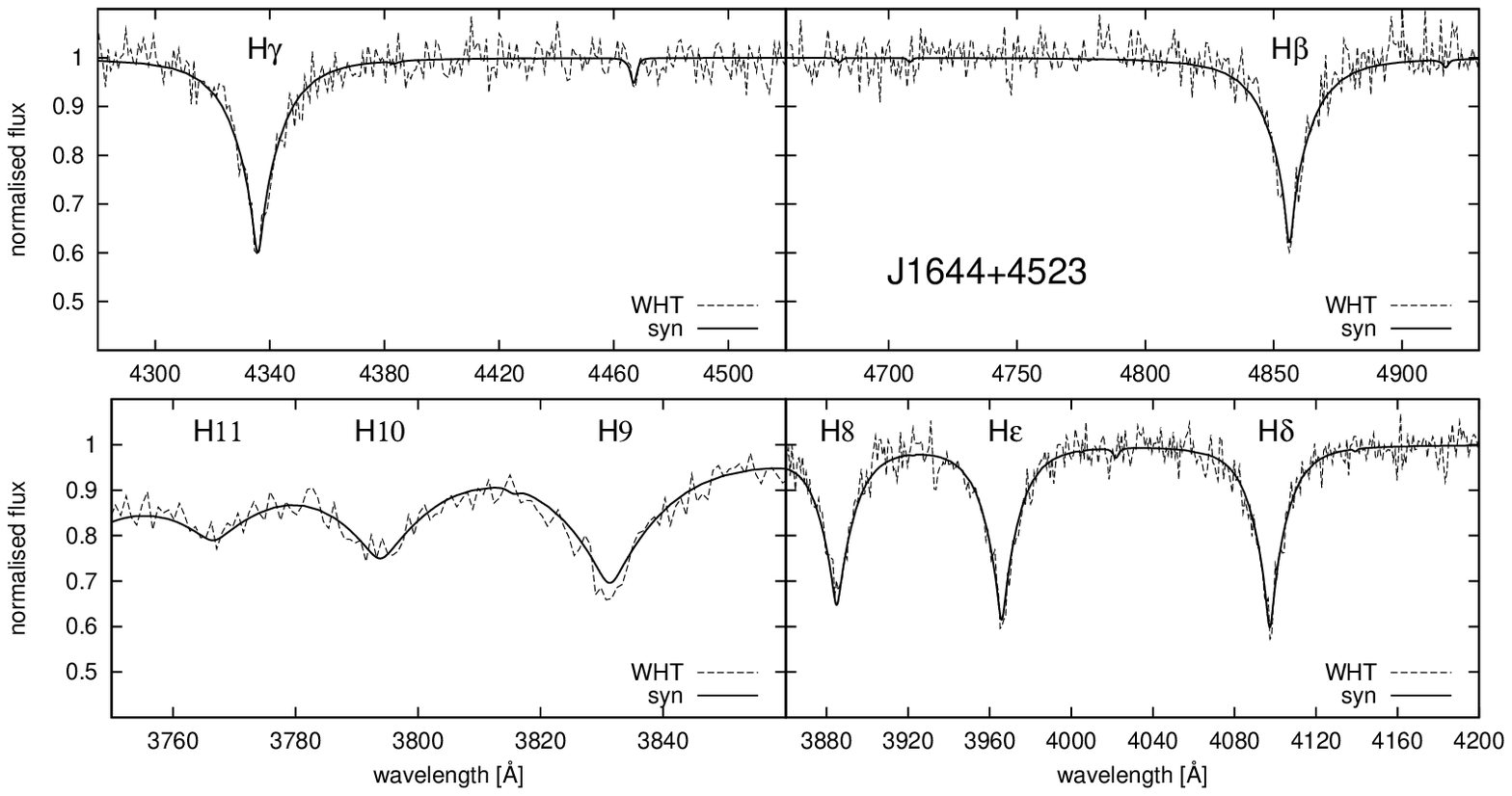}
\includegraphics[angle=0,scale=1.0]{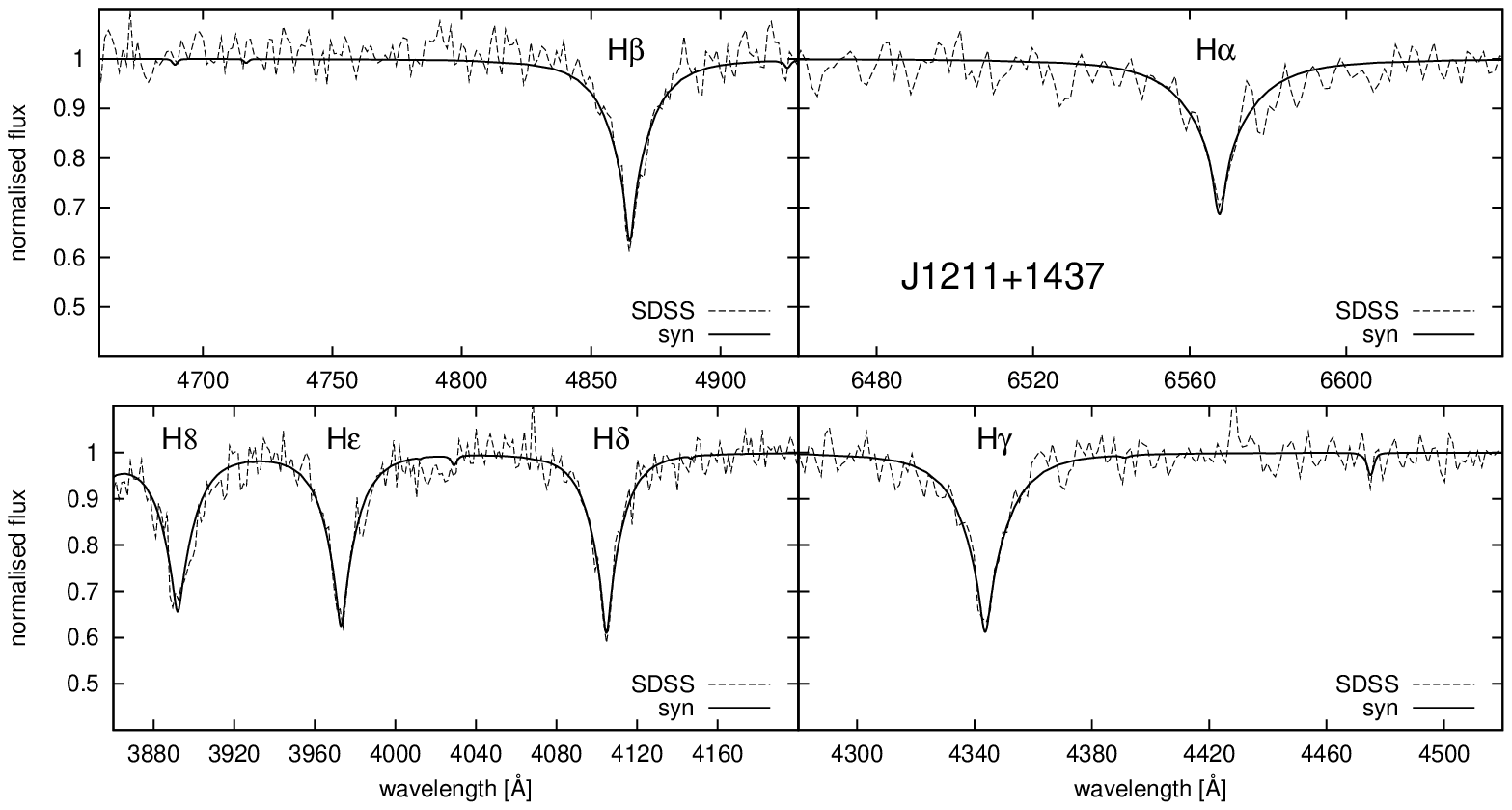}
\caption[Spectral synthesis for the sdBs J1644+4523 and J0948+5516]{Comparison between the synthetic 
model with the observed spectrum (WHT/SDSS) for the sdBs J1644+4523 (top) and J1211+1437 (bottom). 
The overall agreement is good.}
\label{fig:J1644+J1211_spec}
\end{center}
\end{figure*}

In Fig.~\ref{fig:kine} we compare our sdB sample to the white dwarf sample of \cite{2003A&A...400..877P}. 
All sdB stars lie far away from the thin disc population of the white dwarfs. 
They are found in those regions of both diagrams where very few white dwarfs lie; only the rare ones belonging to the halo population. 
The sdB J0845+1352 shows thick disc kinematics (see Fig.~\ref{fig:kine}). 
Nine sdB stars possess halo characteristics, as they reside clearly outside the 3$\sigma$ 
thick disc contour in the diagrams. According to the \textit{e}-$J_{\rm Z}$ diagram, the sdBs can be 
divided into two subgroups. 
Furthermore, the 3D-orbits (see Fig.~\ref{fig:kine2} for examples of an G1 and a G2 star, respectively) are all clearly favour a halo membership. 
However, for 10 stars at least two of three indications are present, which renders them as halo stars, 
following \cite{2006A&A...447..173P}. 
In the \textit{e}-$J_{\rm Z}$ diagram the cluster of subdwarfs with high eccentricities at 
$J_{\rm Z}\approx 0$ clearly catches one's eye. We call this cluster ``group 1'' (G1), as the respective 
stars share similar kinematic properties. 
They have nothing to do with the disc rotation and show only marginal velocity components 
in the direction of the Galactic plane. Such a behaviour is very typical of halo objects, 
as they travel mainly perpendicular through the disc. 
Four subdwarfs have very negative angular momenta, which means that 
they are on retrograde orbits. This group of subdwarfs we term ``group 2'' (G2). 

The two targets with the highest measured PMs turned out to 
be DA white dwarfs. In fact, J1358+4729 was discovered and analysed already by \cite{2006ApJS..167...40E}. They derived atmospheric parameters (\teff\ $=10635\pm61$~K, \logg\ $=8.16\pm0.053$), which are perfectly consistent with our values. 
J1358+4729 shows thick disc kinematics (see Fig.~\ref{fig:kine}) according to both diagrams 
and does not belong to the halo population. The other white dwarf J0849+1455 has not been analysed before, except for a proper motion study by \cite{1992MNRAS.255..521E}, and belongs to the halo population (see Fig.~\ref{fig:kine}). In their last orbits their trajectories never came close to the central part of the Galaxy. 

\subsection{J1211+1437: extreme halo or hyper-velocity star}\label{sec:j1211}


According to the kinematic analysis presented above the sdB star J1211+1437 
 shows the most extreme kinematics in G2 of retrograde orbit stars, 
as it lies far away from the other stars in the kinematic diagrams.

\subsubsection*{J1211+1437 -- a HVS candidate}

The Galactic rest-frame velocity of J1211+1437 is so high (v$_{GRF}=713^{+155}_{-139}$~\kms, see Fig~\ref{fig:J1211_SHIST}) that it exceeds the local Galactic escape velocity of v$_{esc}=507$~\kms\ if we adopt the Galactic potential of \citet{1991RMxAA..22..255A}.   
J1211+1437 is a HVS candidate that could have been ejected from the Galactic centre by the Hills mechanism. In order to test this hypothesis we traced the trajectory back to zero Galactic latitude. 

\begin{figure}[tb]
\begin{center}
 \includegraphics[angle=0,scale=0.75]{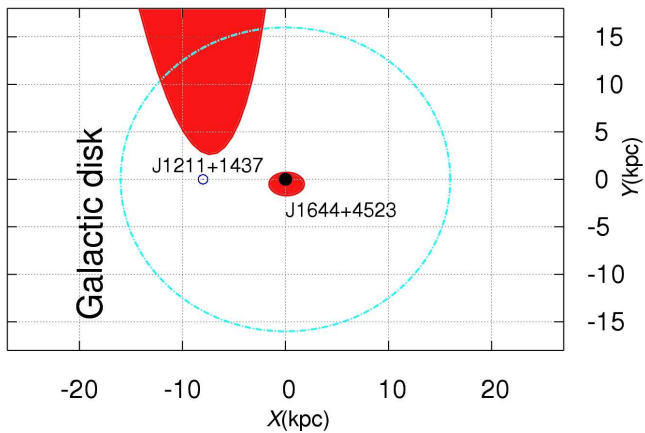}
\caption{Intersection region of the subdwarfs' past trajectories for J1211+1437 and J1644+4523 
with the GD in order to conclude on their origin (grey region). 
For orientation the position of the sun is marked as well.}
\label{fig:CROSSZOOM}
\end{center}
\end{figure}
\begin{figure}
\begin{center}
\includegraphics[scale=0.65]{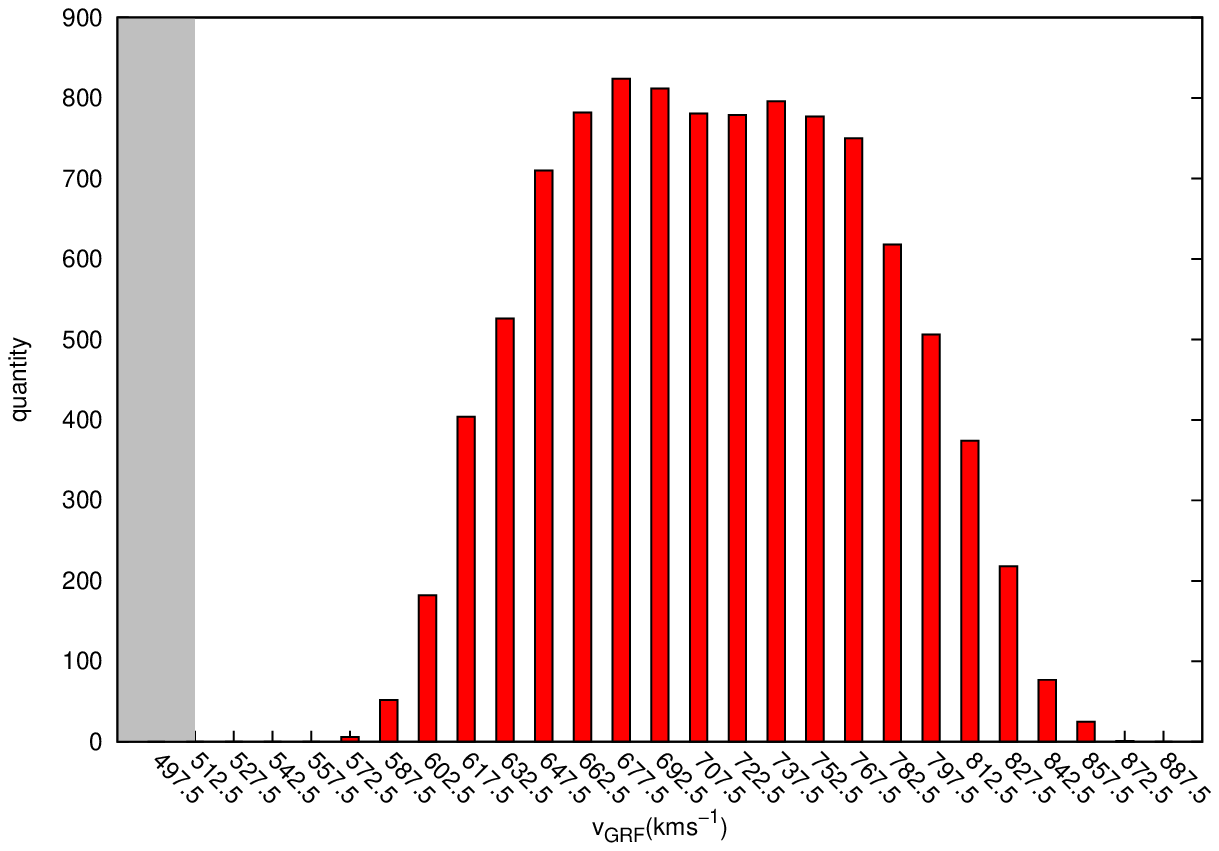}
\caption{
Galactic restframe velocity distribution for J1211+1437 derived from a Monte 
Carlo procedure with a depth of 10,000 \citep[cp. ][]{2009A&A...507L..37T,2010A&A...517A..36T,2010ApJ...718...37P}. The gray shaded area indicate the 
velocities for which the star is bound in the potential of \cite{1991RMxAA..22..255A}. 
}
\label{fig:J1211_SHIST}
\end{center}
\end{figure}

As can be seen from Fig.~\ref{fig:CROSSZOOM} 
J1211+1437 does not originate in the GC. 
Its place of origin is more likely to lie in the Galactic disc, where no SMBH is known to exist. Hence, the \cite{1988Natur.331..687H} 
slingshot mechanism can by excluded. 

The ejection velocity for that star, if it 
would have belonged to a co-rotating disc population, is about $v_{ej}\approx550$~\kms. Hence the star might have been ejected from the 
Galactic disc as a run-away star. 


\subsubsection*{J1211+1437 and the dark matter halo}


If, however, J1211+1437 is an old halo star that experienced several disc passages the star has to be 
bound to the Galaxy. This can only be achieved if the Dark Matter halo is more massive than the adopted one in the potential of \citet{1991RMxAA..22..255A}. 
In general the kinematics of the most 
extreme stars provide the best constraints on the mass of the dark halo \citep{1999MNRAS.310..645W}. 
\cite{2010ApJ...718...37P} found a population II BHB star travelling towards us at such a high speed that a dark matter halo mass of at least $1.7\times 10^{12}$ M$_\odot$ is required to keep the star bound to the Galaxy. This value is considerably larger than adopted by \citet{1991RMxAA..22..255A}. 
Therefore numerical experiments 
were carried out in which the Galactic potential was modified by increasing the mass of the dark matter halo. This constrains the total Galactic mass to exceed
$M_{\rm total}^{\rm new}=2.0\pm^{+2.4}_{-1.2}\times10^{12}$~\msun. The errors are based on the 
Monte Carlo distribution of the space velocity. Although the uncertainties are large, our value is 
perfectly consistent with other proper motion based kinematic mass estimates 
\citep{2010ApJ...718...37P,2003A&A...397..899S,1999MNRAS.310..645W}, which also support a high-mass Galactic halo. 
However, it has to be stressed that this conclusion can only be drawn, if the
orbit of J1211+1437 is bound\footnote{The rest of the sample is bound to the Galaxy, 
although the subdwarfs J1556+4708 and J1644+4523 might exceed the Galactic 
escape velocity if  
their respective errors (see Table~\ref{tab_dist}) are taken into account and the Galactic potential of \citet{1991RMxAA..22..255A} is applied.}. 

\subsection{J1644+4523 - the fastest G1 object}
\begin{figure*}[tb]
\begin{center}
\includegraphics[scale=1.2]{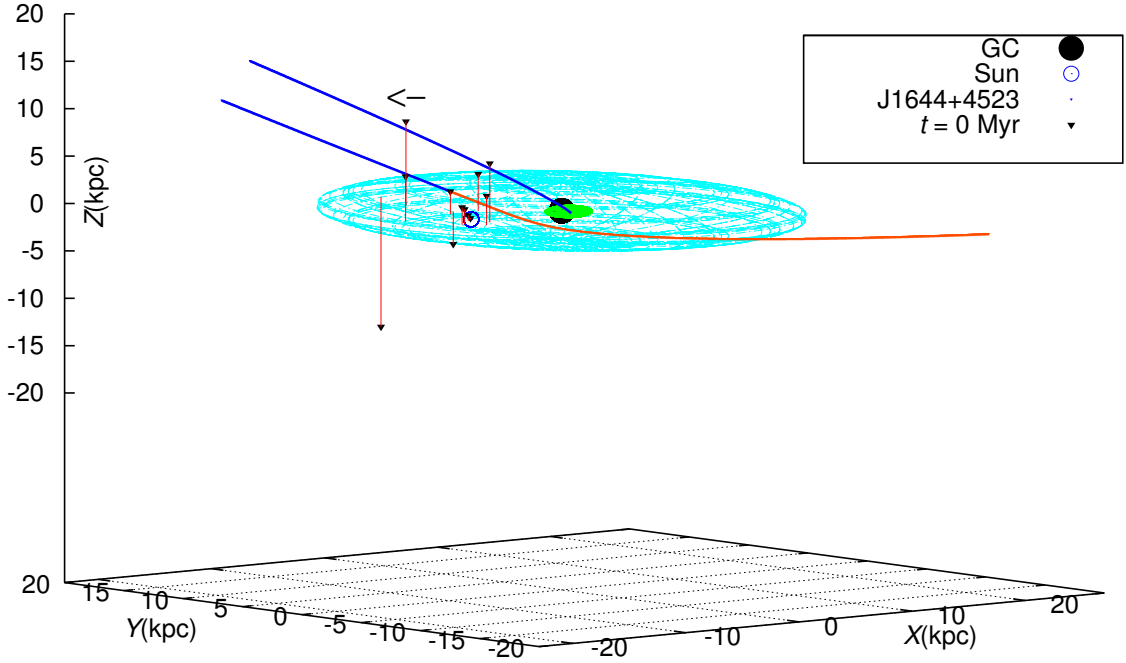}
\caption{3D plot of the current positions of our 12 stars (black triangles) relative to the Galactic Disk (grey), showing also the trajectory for J1644+4523 and the area of potential origin (dark grey).}
\label{fig:J1644_3dplot}
\end{center}
\end{figure*}

Another star that is worth closer inspection is the 
sdB star J1644+4523, which belongs kinematically to 
the G1 group. The star is heading towards us and therefore must be bound. It is the second fastest hot subdwarf of the halo population. 
We measured a GRF velocity of $v_{\rm GRF}=468^{+104}_{-90}$~\kms, which is below the escape velocity of 
$v_{\rm esc}=534$~\kms\ in the Galactic potential of
\cite{1991RMxAA..22..255A}. 
As for J1211+2423 we might consider alternatively that the star was ejected as a run-away star.
We calculated the trajectory backwards into the past and found that the star might have originated 
in the Galactic bulge (GB), which can be seen from Fig.~\ref{fig:J1644_3dplot}. 
The last pericenter passage occurred at a distance of only $\sim0.5\pm1.0$~kpc from the 
Galactic centre (see Fig.~\ref{fig:CROSSZOOM}) and the apocenter distance of the star's trajectory is located far out in the halo
at $\sim115$~kpc. 
The high ejection velocity of $v_{\rm ej}=597$~\kms\ could possibly 
be consistent with the ejection by the SMBH in the GC, as the error ellipse 
encloses the GC (see Fig.~\ref{fig:CROSSZOOM}). 
The time-of-flight (TOF) is much longer ($TOF=1.27$~Gyr) than the helium core burning phase lasts ($\approx200$~Myr). This means that J1644+4523 evolved to a subdwarf a 
long time after it was possibly ejected from the GB. 
This is very problematic, as EHB stars most likely require a binary interaction scenario to form \citep{2003MNRAS.341..669H}. In this case the ejection by the SMBH would disrupt any initial binary 
and the sdB cannot form. One solution could be a shorter TOF. 
The kinematic analysis of J1211+2423 indicated that the dark matter halo may be more massive than assumed in the Galactic potential of \cite{1991RMxAA..22..255A}. Therefore we repeated the kinematical calculations for a higher dark halo mass and adopted $M_{\rm halo}^{\rm new}=3.4\times10^{12}$~\msun\ as 
suggested by \cite{2009ApJ...691L..63A} with a mass distribution out to 100~kpc, following 
\cite{1991RMxAA..22..255A}. The place of the last disc intersection remained almost unchanged but
the time-of-flight is strongly reduced to $TOF=145$~Myr, which is consistent with the lifetime of hot subdwarfs. 
Hence, in a high-mass halo the star could have been ejected as subdwarf and did not have to evolve after the ejection. In that case the last pericenter passage occurred at a slightly higher 
distance of $\sim0.65$~kpc from the Galactic centre and 
the apocenter distance of the star’s trajectory is located in the halo at only $\sim22$~kpc. 

\subsection{The entire sdB sample}
Hence we consider it worthwhile to inspect the run-away scenario for all target sdB stars by taking a closer look at their trajectories from the Galactic plane to their present location in the Galactic halo. 
Fig.~\ref{fig:CROSS} displays the areas in the plane from which the stars could have originated for the 
kinematic group G1 (top) and G2 (bottom).  As can be seen the G1 stars would originate from the inner Galactic disc or bulge, in particular J1644+4523, whereas the G2 stars (except J1632+2051) would come from the outer disc\footnote{The remaining thick disc subdwarf J0845+1352 shows a trajectory, which intersects 
within the Galactic disc.}.
The case of J1211+1437 demonstrates that some of our targets could possibly be 
ejected run-away stars. 
If so the places of plane crossings of the G1 stars would not rule out an origin in the Galactic centre (except for J0120+0137). However, a Galactic centre 
origin of the G2 stars is excluded as their potential places of origin do not include the central part of the Galaxy. 

\subsection{Kinematics of hot subdwarfs}

Up to now, the only 3D kinematical study of such sdBs in the field was 
performed by \cite{2004A&A...414..181A}. 
From a sample of 114 sdBs they found 15 to have orbits, which 
differ considerably from disc orbits, while the vast majority is consistent with the disc population. 
 \cite{2004A&A...414..181A} noticed that their sample may contain close binary stars that are radial velocity variable, because the fraction is known to be about 40\%. At the time of writing several stars of their sample had not been checked for radial velocity variability, but have been shown to be RV variable later on. We do not correct for this effect.
 
\begin{figure}[tb]
\includegraphics[scale=0.72]{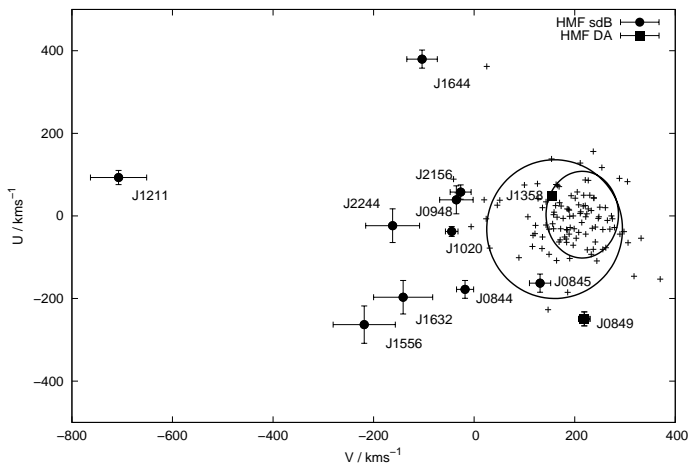}
\caption{   
Kinematic V-U diagram for the programme sdB stars in comparison with the sample (+ signs) of \cite{2004A&A...414..181A}. The solid ellipses render the $3\sigma$-thin and thick disc contours of \cite{2006A&A...447..173P}.
}
\label{fig:altmann} 
\end{figure}

\begin{figure*}[tb]
\begin{center}
\includegraphics[scale=1.3,angle=0]{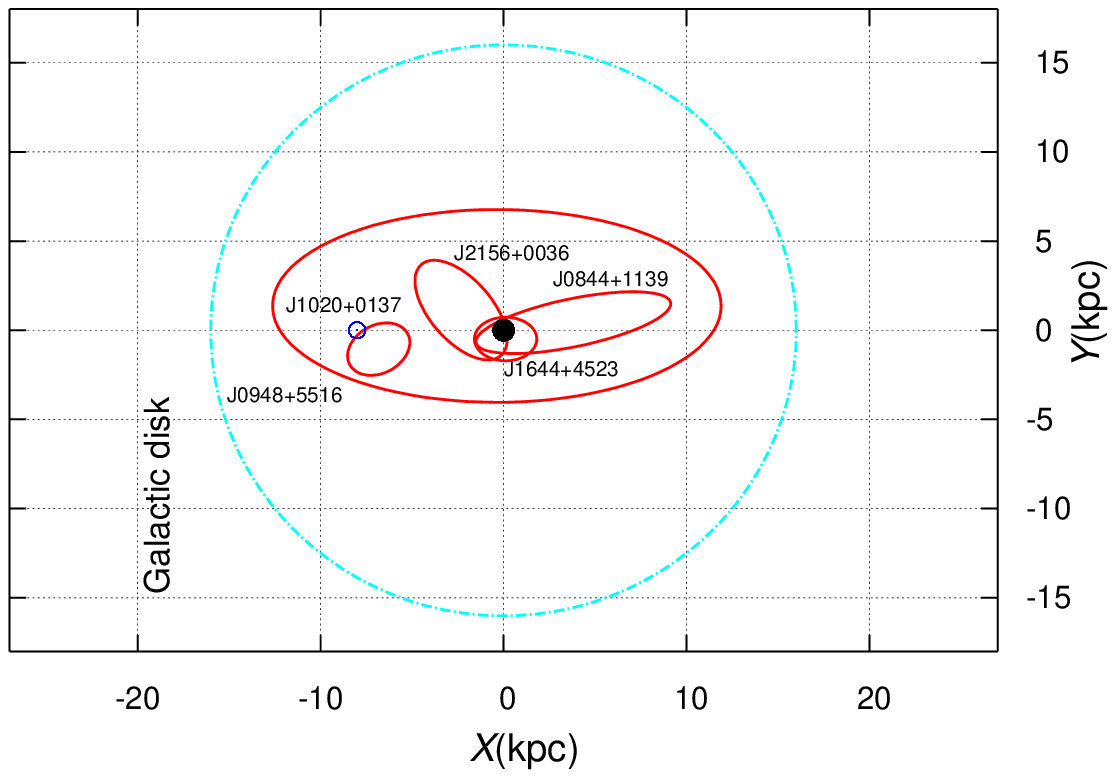}
\includegraphics[scale=1.3,angle=0]{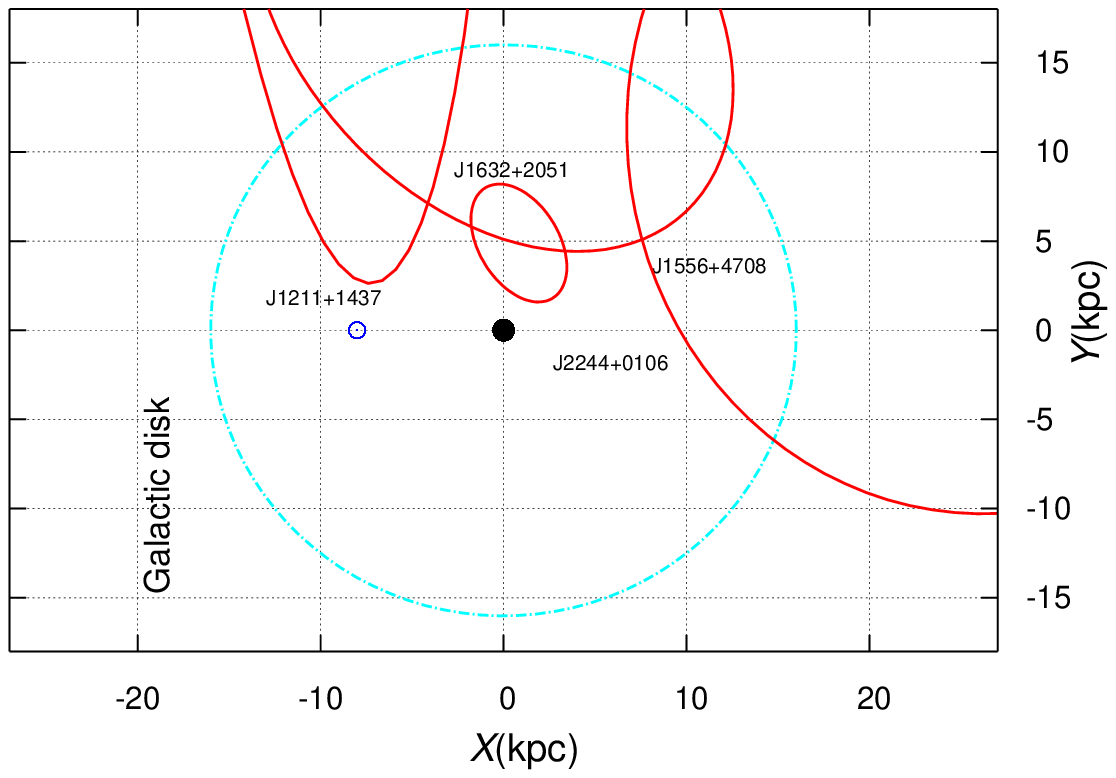}
\caption{Intersection region of the halo subdwarfs' past trajectories with the GD in order to conclude 
on their origin. Note that in the group G1, 4 out of 5 subdwarfs are consistent with a GC origin 
(\textit{top}). 
In the extreme group G2 we found 3 subdwarfs, which might originate in the outer 
Galactic rim (\textit{bottom}) and have been likely produced by supernova runaway ejection. 
For orientation the position of the sun is marked as well.}
\label{fig:CROSS}
\end{center}
\end{figure*}

 We compare the kinematical properties in the $V$-$U$-plane of their sample to ours in Fig.~\ref{fig:altmann}.
 Applying the criteria of \citet{2006A&A...447..173P} the vast majority of the 
  \citet{2004A&A...414..181A} sample belong to the thin disc, with about two dozen thick disc and a dozen halo stars. The most eye-catching difference between the two samples in Fig.~\ref{fig:altmann} is the absence of stars with rapid retrograde orbits, i.e. our class G2. Only very few halo sdBs in the sample of \citet{2004A&A...414..181A} would be classified as G1 (showing essentially no Galactic rotation). 
However, our sample lacks stars with large positive $V$ velocities. 

When comparing the two samples we have to consider that the
stars in the sample of \citet{2004A&A...414..181A} are much brighter and, hence, nearer by than those in our sample. Consequently, we probe a much larger volume of space. 

\section{Summary and conclusion}\label{sec:conclu}

Thirty-nine high-velocity faint blue stars have been identified from the MUCHFUSS survey. We were able to measure proper motions significantly different from zero for 12 of them by determining positions of 60 years time base. Quantitative spectral analyses were performed to derive their distances. From an investigation of their kinematics 
nine sdB stars were identified as halo members and one as belonging to the thick disc. The two remaining stars turned out to be white dwarfs in the halo and the thick disc, respectively.
Two distinctive kinematic groups among the sdB stars emerged: the normal halo subdwarfs (G1) with low Galactic rotation and 
the extreme halo subdwarfs (G2) on highly-eccentric retrograde orbits.

The extreme halo star J1211+1437 is of particular interest as it would be unbound to the Galaxy if the standard Galactic potential was applied. Hence it would be the second HVS among the hot subdwarfs and the first one of spectral type sdB. However, assuming the star is a population II object bound to the Galaxy, we can derive a lower limit to the mass of the dark matter halo of $2\times 10^{12} M_\odot$. Other stars, J1644+4523 in particular, that are approaching the Earth have to be bound to the Galaxy and therefore might provide constraints on the dark matter mass. 

The existence of two kinematically distinct groups indicates different origins. In the ejection scenario the class G1 stars would originate from the inner Galaxy, whereas the G2 ones would come from its outer parts.   
However, if some of the sdB stars in our sample were run-away stars, we have to find an ejection mechanism. The G1 stars (except J0120+0137) may originate from the Galactic centre and the SMBH slingshot mechanism might work for them. 
For the G2 stars an origin in the Galactic centre can be excluded. Hence we have to find another mechanism to explain their origin. 

Kinematic studies among massive B-type stars indicate that variations of the typical 
runaway-scenario might provide an answer \citep{2008A&A...483L..21H,2010ApJ...711..138I}. 
Conveniently, sdBs most likely require a binary interaction scenario to form \citep{2003MNRAS.341..669H}. 
In principle one can think of a variation of the RLOF mechanism as suggested by \cite{1998A&A...335L..85N} 
in the context of the formation of undermassive white dwarfs. 
If the subdwarf's RGB progenitor is losing its envelope to a massive white dwarf, an asymmetric accretion induced collapse could occur. The system can then be disrupted, with the accretor leaving the system as a high-velocity neutron star. As computed by \cite{2000ApJS..128..615M}, the companions SN explosion does not have to happen at the tip of the RGB for the envelope to be lost and the core to end up as an sdB star. In their calculations, the SN explosion itself can strip 96 to 98\%\ of the envelope from an RGB star, effectively leaving a naked He core. If the core is massive enough for helium burning it can experience a late core He flash, and end up as a single EHB star. As pointed out by 
\cite{2009CoAst.159...75O}, in both these cases the abandoned subdwarf would end up single and in an unusual galactic orbit. 

Recently, \cite{2009A&A...508L..27W} and \cite{2009A&A...493.1081J} suggested a single degenerate SN Ia scenario to explain the formation 
of the helium sdO US~708. Accordingly, a white dwarf accretes matter from am helium star in a close binary system. After exceeding the Chandrasekhar mass-limit, the white dwarf explodes as a type Ia supernova. The orbital period has shrunk to about 1h and the corresponding orbital velocity 
of the hot He-sdO companion may exceed 500~\kms. After the explosion of the primary, the binary is 
disrupted and the sdO companion is released at its orbital velocity. In addition it may be speeded up by the SN kick.
However, all our programme stars are sdB stars, which retain a hydrogen-rich envelope of 10$^{-4}$ to 10$^{-2}$ M$_\odot$. Hence, it appears unlikely that the scenario applies to high-velocity sdB stars. 
  
As some globular clusters are known to host sdB stars, ejection from such 
clusters may give rise to high-velocity sdB stars. Alternatively, the programme stars may stem from the disruption of a satellite galaxy in the Galactic halo.

We shall extend our study as the MUCHFUSS survey proceeds. As demonstrated, the high-velocity hot subdwarfs are important tools to constrain the mass of the 
Galactic dark matter halo. Its full potential will develop once the GAIA mission will provide much more accurate distances and proper motions.

\begin{acknowledgements}
A.T., S.G. and H.H. are supported by the Deutsche Forschungsgemeinschaft (DFG) through grants HE1356/45-1, HE1356/49-1, and HE1356/44-1, respectively. Travel to the DSAZ (Calar Alto, Spain) was supported by DFG under grants HE1356/48-1 and HE1356/50-1. \newline
R.H.{\O} has received funding from the European Research Council under the European Community's Seventh Framework Programme
(FP7/2007--2013)/ERC grant agreement N$^{\underline{\mathrm o}}$\,227224
({\sc prosperity}), as well as from the Research Council of K.U.Leuven grant agreement GOA/2008/04. \newline
We thank Detlev Koester, who provided the models for the white dwarfs. 
R.-D.S. thanks Doug Finkbeiner for his
help accessing the SDSS data at Princeton University. \newline
Funding for the SDSS and SDSS-II has been provided by the Alfred P.
Sloan Foundation, the Participating Institutions, the National Science
Foundation, the U.S. Department of Energy, the National Aeronautics
and Space Administration, the Japanese Monbukagakusho, the Max Planck
Society, and the Higher Education Funding Council for England. The
SDSS Web Site is http://www.sdss.org/.
\end{acknowledgements}

\bibliography{references}
\bibliographystyle{aa}

\end{document}

%% file: tab_RV2.list
SDSSJ084447.00+113910.0 &  J0844+1139 &   sdB    &    SDSS  &   2453801.1251331      &   +202.7 \pmp   5.0   & 52 \\
                        &             &          &    NTT   &   2454755.86312183     &   +219.5 \pmp  30.4   & 13 \\
SDSSJ084556.85+135211.3 &  J0845+1352 &   sdB    &    SDSS  &   2453799.765353       &   +115.1 \pmp   9.6   & 33 \\
                        &             &          &    NTT   &   2454756.86725607     &   + 92.0 \pmp  15.1   & 22 \\
SDSSJ094850.47+551631.7 &  J0948+5516 &   sdB    &    SDSS  &   2451991.6335069      &   -138.3 \pmp  12.2   & 17 \\
                        &             &          &    WHT   &   2454587.46586414     &   -158.5 \pmp  17.8   & 9  \\
                        &             &          &    WHT   &   2454588.38885923     &   -127.4 \pmp   6.6   & 15 \\
SDSSJ102057.16+013751.3 &  J1020+0137 &   sdB    &    SDSS  &   2451997.186412       &   +243.3 \pmp   8.3   & 48 \\
                        &             &          &    FORS  &   2454638.47970756     &   +239.4 \pmp   7.8   & 36 \\
SDSSJ121150.27+143716.2 &  J1211+1437 &   sdB    &    SDSS  &   2453466.7114641      &   +220.2 \pmp   9.5   & 32 \\
                        &             &          &    FORS  &   2454575.76860445     &   +207.5 \pmp  25.6   & 11 \\
SDSSJ155635.80+470851.8 &  J1556+4708 &   sdB    &    SDSS  &   2452354.493235       &   -393.5 \pmp  19.9   & 18 \\
                        &             &          &    WHT   &   2454586.60330613     &   -385.4 \pmp   7.4   & 14 \\
SDSSJ163213.05+205124.0 &  J1632+2051 &   sdB    &    SDSS  &   2453224.6422685      &   -238.6 \pmp   6.7   & 38 \\
                        &             &          &    FORS  &   2454656.62186841     &   -237.9 \pmp   2.8   & 45 \\
SDSSJ164419.44+452326.7 &  J1644+4523 &   sdB    &    SDSS  &   2452051.9386574      &   -314.1 \pmp   4.5   & 32 \\
                        &             &          &    WHT   &   2454586.71567501     &   -312.8 \pmp  11.6   & 31\\
SDSSJ215648.71+003620.7 &  J2156+0036 &   sdB    &    SDSS  &   2452933.5838194      &   -177.2 \pmp   6.5   & 32 \\
                        &             &          &    NTT   &   2454757.70342858     &   -190.1 \pmp  37.2   & 13 \\
SDSSJ224451.81+010630.9 &  J2244+0106 &   sdB    &    SDSS  &   2452146.8122801      &   -166.0 \pmp   5.9   & 24 \\
                        &             &          &    NTT   &   2454756.61349441     &   -149.1 \pmp  19.2   & 17 \\
SDSSJ084938.85+145503.2 &  J0849+1455 &   DA     &    SDSS  &   2453818.6548958      &   +119.0 \pmp  37.2   & 15 \\
                        &             &          &    NTT   &   2454757.87125523     &   + 82.6 \pmp  33.9   & 9  \\
SDSSJ135824.61+472931.6 &  J1358+4729 &   DA     &    SDSS  &   2452723.9695428      &   -260.6 \pmp  14.2   & 24 \\
                        &             &          &    WHT   &   2454586.54874204     &   -257.0 \pmp  14.3   & 22 \\

%% file: tab_PMok.dat
J0844+1139 & 16.11 \pmp 0.02 &  -11.5 \pmp  2.6  &  -11.9  \pmp  2.1 &   6/ 7 \\
J0845+1352 & 17.42 \pmp 0.02 &  -10.7 \pmp  2.3  &   -3.7  \pmp  2.6 &   5/14 \\
J0948+5516 & 18.59 \pmp 0.03 &   -1.9 \pmp  2.9  &   -7.4  \pmp  1.7 &   7/12 \\
J1020+0137 & 16.91 \pmp 0.02 &   -2.4 \pmp  1.3  &   -7.3  \pmp  1.2 &  12/15 \\
J1211+1437 & 17.87 \pmp 0.03 &  -12.1 \pmp  1.8  &  -27.2  \pmp  1.4 &  13/14 \\
J1556+4708 & 18.61 \pmp 0.03 &   -6.3 \pmp  1.9  &   +1.3  \pmp  1.8 &  10/14 \\
J1632+2051 & 17.62 \pmp 0.03 &  -12.5 \pmp  3.0  &   -1.6  \pmp  3.6 &   7/16 \\
J1644+4523 & 17.36 \pmp 0.02 &   +4.7 \pmp  2.8  &  -26.1  \pmp  3.3 &   8/13 \\
J2156+0036 & 17.91 \pmp 0.03 &   -1.3 \pmp  1.6  &   -7.4  \pmp  1.2 &  26/13 \\
J2244+0106 & 18.14 \pmp 0.02 &   +2.1 \pmp  0.8  &   -3.7  \pmp  1.3 &  30/14 \\
J0849+1455 & 18.85 \pmp 0.03 &  -49.5 \pmp  1.8  &   +6.4  \pmp  1.6 &   8/ 9 \\
J1358+4729 & 18.03 \pmp 0.03 &  +76.1 \pmp  2.4  &  +14.4  \pmp  2.4 &  14/15 \\

%% file: PM_ok_paramtex.tab_RD
J0844+1139   &  sdB      &     28600 \pmp  740 &  5.33 \pmp 0.12  &  $-$2.6 \pmp 0.4 \\             
J0845+1352   &  sdB      &     24570 \pmp 1070 &  5.59 \pmp 0.17  &  $<$-3.0        \\             
J0948+5516   &  sdB      &     33890 \pmp  560 &  5.84 \pmp 0.20  &  $-$1.5 \pmp 0.2 \\    
J1020+0137   &  sdB      &     28960 \pmp  760 &  5.39 \pmp 0.14  &  $<$-3.0        \\    
J1211+1437   &  sdB      &     32530 \pmp  580 &  5.52 \pmp 0.16  &  $-$2.9 \pmp 0.5 \\    
J1556+4708   &  sdB      &     34410 \pmp 1280 &  5.34 \pmp 0.25  & $<$-3.0        \\             
J1632+2051   &  sdB      &     26870 \pmp  610 &  5.31 \pmp 0.09  &  $-$2.1 \pmp 0.2 \\    
J1644+4523   &  sdB      &     31680 \pmp  410 &  5.78 \pmp 0.11  &  $-$2.9 \pmp 0.3 \\    
J2156+0036   &  sdB      &     29080 \pmp  810 &  5.60 \pmp 0.15  &  $-$2.5 \pmp 0.3 \\    
J2244+0106   &  sdB      &     33580 \pmp  680 &  4.75 \pmp 0.20  &  $-$1.5 \pmp 0.2 \\
J0849+1455   &  DA       &     27750 \pmp  830 &  7.32 \pmp 0.15  &    -        \\    
J1358+4729   &  DA       &     10640 \pmp  110 &  8.10 \pmp 0.07  &    -        \\             

%% file: tab_vGRF.list
J0844+1139 &  3.14 \pmp 0.26 &  202.7 \pmp  5.0 & $202^{+55}_{-46}$   & 513 \vspace{0.5mm}  \\
J0845+1352 &  3.50 \pmp 0.38 &  115.1 \pmp  9.6 & $230^{+70}_{-61}$   & 511 \vspace{0.5mm}  \\
J0948+5516 &  6.51 \pmp 0.88 & -142.1 \pmp  8.6 & $106^{+78}_{-35}$   & 490 \vspace{0.5mm}  \\
J1020+0137 &  4.31 \pmp 0.39 &  238.0 \pmp  8.3 & $107^{+20}_{-13}$   & 513 \vspace{0.5mm}  \\
J1211+1437 &  6.45 \pmp 0.68 &  220.1 \pmp  9.5 & $713^{+155}_{-139}$ & 507 \vspace{0.5mm}  \\
J1556+4708 & 11.76 \pmp 2.03 & -385.4 \pmp  7.4 & $353^{+174}_{-106}$ & 487 \vspace{0.5mm}  \\
J1632+2051 &  6.09 \pmp 0.36 & -236.1 \pmp  3.2 & $275^{+120}_{-110}$ & 544 \vspace{0.5mm}  \\
J1644+4523 &  3.68 \pmp 0.27 & -314.1 \pmp  4.5 & $468^{+104}_{-90}$  & 534 \vspace{0.5mm}  \\
J2156+0036 &  5.37 \pmp 0.54 & -177.2 \pmp  6.5 & $85^{+58}_{-44}$    & 534 \vspace{0.5mm}  \\
J2244+0106 & 18.29 \pmp 2.45 & -166.0 \pmp  5.8 & $200^{+186}_{-156}$ & 458 \vspace{0.5mm}  \\
J0849+1455 &  1.11 \pmp 0.11 &  119.0 \pmp 37.2 & $354^{+33}_{-34}$   & 531 \vspace{0.5mm}  \\
J1358+4729 &  0.14 \pmp 0.01 & -260.6 \pmp 14.2 & $295^{+10}_{-11}$   & 541 \vspace{0.5mm}  \\